# Benchmarking MD systems simulations on the Graphics Processing Unit and Multi-Core Systems

*Iuliana Marin[1], Nicolae Goga[1,2], Maria Goga[3]*

[1]Faculty of Engineering in Foreign Languages,
University POLITEHNICA of Bucharest, Romania

[2]Molecular Dynamics Group
University of Groningen, Netherlands

[3]Technical University of Construction,
Bucharest, Romania

marin.iulliana25@gmail.com, n.goga@rug.nl,
goga.maria@yahoo.com

*Abstract*—Molecular dynamics facilitates the simulation of a complex system to be analyzed at molecular and atomic levels. Simulations can last a long period of time, even months. Due to this cause the graphics processing units (GPUs) and multi-core systems are used as solutions to overcome this impediment. The current paper describes a comparison done between these two kinds of systems. The first system used implies the graphics processing unit, respectively CUDA with the OpenMM molecular dynamics package and OpenCL that allows the kernels to run on the GPU. This simulation is done on a new thermostat which mixes the Berendsen thermostat with the Langevin dynamics. The second comprises the molecular dynamics simulation and energy minimization package GROMACS which is based on a parallelization through MPI (Message Passing Interface) on multi-core systems. The second simulation uses another new thermostat algorithm related respectively, dissipative particle dynamics - isotropic type (DPD-ISO). Both thermostats are innovative, based on a new theory developed by us. Results show that parallelization on multi-core systems has a performance up to 33 times greater than the one performed on the graphics processing unit. In both cases temperature of the system was maintained close to the one taken as reference. For the simulation using the CUDA GPU, the faster runtime was obtained when the number of processors was equal to four, the simulation speed being 3.67 times faster compared to the case of only one processor.

I. INTRODUCTION

Molecular dynamics (MD) allows the simulation of complex systems to be performed on computers at molecular level by applying Newtonian equations of motion and/or constraints. Such simulations allow acquiring information about the thermodynamic and kinetic properties of a molecular system. Simulations of molecular systems on a timescale of milliseconds analyze the dynamics of large groups of molecules. Sometimes coarse-grained molecules are used. Coarse-grained molecules consist of meta-particles corresponding to group up of atoms. that In these atomic systems quantum mechanical computations are applied. The interactions between the atoms are the starting point for the MD computations at each simulation step.

Molecular dynamics simulations provide detailed information about the molecular and atomic levels which can be used in drug discovery [1]. The illness progression can be controlled by adjusting the activity of a proteins. Drug to drug interactions can be studied using this approach. A successful example is that of Raltegravir which is an anti-HIV-AIDS medicine that resulted from a dynamics simulation [2,3]. The structural information of G-protein coupled receptors (GPCRs) are introduced and studied with the help of molecular dynamics simulations [4].

Some other examples of MD uses: about 18 million small components were tested using MD simulations against the dengue virus, a severe death cause of the tropical and sub-tropical regions [5]. Using the same technique, benzoxazepine moiety proved to be a mechanistic target of rapamycin (mTOR) inhibitor for the medical care of lung cancer [6]. mTOR regulates the growth of cells.

Another benefit is that not only illnesses and drug interactions can be examined, but also the strength of materials. Through MD simulations it can be studied the impact of titanium and magnesium nanoinclusions on the strength of aluminum [7]. Electrical conductance can also be tested by applying MD simulations. For example, an aqueous pore from a double layer of lipids proves to have electrical conductance, based on the continuum theories of Poisson, Nernst and Planck [8].

A very important factor that needs to be taken into consideration is the time needed to simulate such systems, up to weeks and even months. One solution for this problem is to introduce parallelization on different types of hardware systems. By applying a set of processes and tools which use the graphics processing unit or parallelization on multi-core systems, the overall simulation time can be reduced. The current paper describes a comparison done between those two kinds of systems: graphics processing units (GPUs) and multi-core systems.

The article is organized as follows. In the next section, the basic theory related to molecular dynamics is presented, introducing notions about stochastic dynamics, the Berendsen thermostat, Maxwellian velocities and the dissipative particle dynamics. Section 3 outlines a comparison done between the simulations performed on the graphics processing unit and the multi-core systems. The presented simulation on graphics processing unit regards the system composed of the OpenMM molecular dynamics package which uses: CUDA, GPU and OpenCL. The simulation on multi-core systems comprises the molecular dynamics simulations package GROMACS (that in early versions worked with OpenMM), and MPI (Message Passing Interface) communication. Section 4 summarizes the conclusions.

## II. MOLECULAR DYNAMICS THEORY

All the atoms within a simulation system are described by forces, positions and velocities. In a 3D simulation environment of atoms, the number of degrees of freedom $n$ is equal to three times the number of particles $N$. Each simulation step they get updated according to the following formula:

$$v_i = v_{old} + \frac{F}{m}\Delta t \qquad (1)$$

$$x_i = x_{old} + v_i \Delta t \qquad (2)$$

The kinetic energy is expressed as

$$E_c = \frac{nk_B T_{ref}}{2} \qquad (3)$$

where the Boltzmann constant $k_B$ is equal to $1.38 * 10^{-23}$ J/K.

Stochastic dynamics, also known as velocity Langevin dynamics augment Newton's equations of motion with a friction and noise factor [9].

The velocity of particles is given by

$$v = f * v_m + R * \varepsilon \qquad (4)$$

where $f$ represents friction and it is approximately 1, $R$ is a constant, $\varepsilon$ is the white noise, while $v_m$ is expressed as

$$v_m = v\left(t - \frac{\Delta t}{2}\right) + \frac{F}{m}\Delta t \qquad (5)$$

where $R$ results from

$$R = \sqrt{(1 - f^2) * \frac{k_B T_{ref}}{m}} \qquad (6)$$

Velocity can be rewritten as

$$v = f * v_m + \sqrt{(1 - f^2) * \frac{k_B T_{ref}}{m}} * \varepsilon \qquad (7)$$

The coordinates and the velocities of atoms are updated using:

$$\begin{cases} x(t + \Delta t) = x(t) + \Delta t * \frac{\Delta v + v_m}{2} \\ v\left(t + \frac{\Delta t}{2}\right) = \Delta v + v \end{cases} \qquad (8)$$

Thermostats are used because a system should be maintained at a specific temperature, namely reference temperature. Bellow we will discuss two types of thermostats: Berendsen and DPD thermostats.

The **Berendsen thermostat** applies a weak coupling methods to an external bath [10]. The deviations of the current temperature $T$ according to the reference temperature are improved by multiplying a factor $\lambda$ to the atom velocities. The system temperature is ameliorated, the deviation decaying exponentially with a time constant $\tau$.

$$\frac{dT}{dt} = \frac{1}{\tau}(T_0 - T) \qquad (9)$$

The temperature will be adjusted using

$$T = T_0 - C\, e^{-t/\tau} \qquad (10)$$

where C represents the heat capacity.

The scaling factor is computed by applying

$$\lambda^2 = 1 + \frac{\Delta t}{\tau}\left(\frac{T_0}{T} - 1\right) \qquad (11)$$

In the case when $m$ constraints are applied inside the system, the degrees of freedom is equal to three times the difference between $N$ and $m$. The correction of the lost degrees of freedom is done by applying Maxwellian velocities for all the particles which have the distance $d_i$ between them by applying the velocity vector projection $p_i$ upon the tangent plane at the contact with the hypersurface.

$$v_m^i = k_B T(d_i - p_i)\frac{1}{m_i} \qquad (12)$$

The velocity is computed according to the constraints that occur as

$$v_c^i = (d_i - p_i)v_m^i \qquad (13)$$

The final velocity is equal to

$$v_i^c = v_i + v_m^i - v_c^i \qquad (14)$$

In the combined, new, thermostat developed by us the velocities $v_m$ are scaled with the Berendsen scaling factor $\lambda$ and they are the input of equation (4). Then the Langevin dynamics is applied. In this way both advantages of Berendsen and Langevin dynamics can be used.

**DPD (Dissipative Particle Dynamics)** applies friction and noise to particle pairs, being a thermostat that maintains the total momentum of a molecular system. For the isotropic type (DPD-ISO) the friction direction is the same as the one of the velocity. The velocity noise factor is given by

$$g = \sqrt{f(2-f)k_B \frac{T_{ref}}{\mu}} \qquad (15)$$

where the reduced mass of the two particles $\mu$ is

$$\mu = \frac{m_i m_j}{m_i + m_j} \quad (16)$$

The relative velocity vector of two atoms, A and B, is computed as

$$v = v_A - v_B \quad (17)$$

Three random numbers ε ($\varepsilon_1$, $\varepsilon_2$, $\varepsilon_3$) are selected from a standard normal distribution. The vector Δv is

$$\Delta v = -f * v + g * \varepsilon \quad (18)$$

The values of the velocities are renewed by performing

$$\begin{cases} v_A = v_A + \Delta v \frac{m_B}{m_A + m_B} \\ v_B = v_B + \Delta v \frac{m_A}{m_A + m_B} \end{cases} \quad (19)$$

The coordinates are updated for the 3 * (N - m) constraints by setting

$$x'_i = x_i^c + (v'_i - v_i^c) * \frac{\Delta t}{2} \quad (20)$$

and also the velocities as follows

$$v_i\left(t + \frac{\Delta t}{2}\right) = v_i^c + \frac{2 * \left[x_i\left(t + \frac{\Delta t}{2}\right) - x'_i\right]}{\Delta t} \quad (21)$$

### III. COMPARISON BETWEEN SIMULATIONS PERFORMED ON THE GRAPHICS PROCESSING UNIT AND MULTI-CORE SYSTEMS

#### A. Simulation on the graphics processing unit

The simulation performed on the graphics processing unit (GPU) consisted of: 1) the OpenMM molecular dynamics package (that works in early versions of Gromacs together with it) that uses 2) CUDA GPU, together with OpenCL for a 3) molecular model of 2269 Villin protein and water solvent atoms upon which we applied our 4) new stochastic dynamics combined with the Berendsen thermostat [11]. OpenCL allows the kernels to run on GPUs.

As illustrated in Fig. 1, the GPU has as input data: the coordinates, velocities and the Langevin integrator. The code of the integrator is specified in the kernel function that triggers several threads. The simulation is done both on the CPU and on the GPU in parallel. After each iteration the coordinates and the velocities of atoms will be renewed.

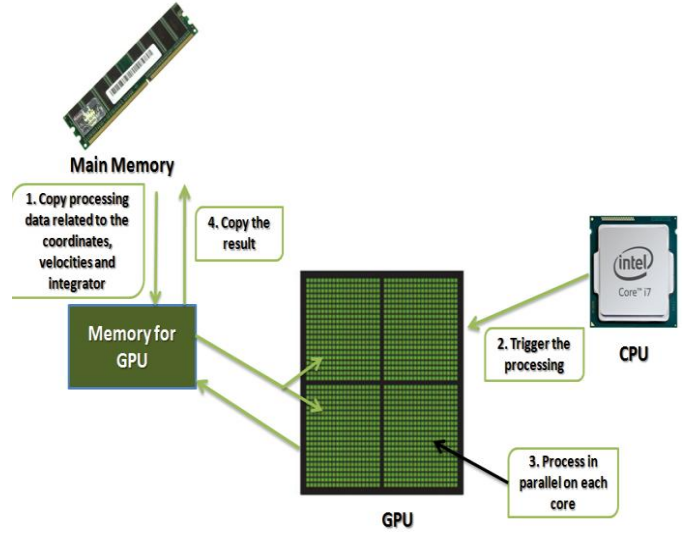

Fig. 1. CUDA processing flow

For all the atoms within the system the kinetic energies and the sum of attraction and repulsion forces per particle are computed. Their velocity, position and kinetic energy are updated according to the new stochastic dynamics (1-3), after which the Berendsen thermostat is applied (9-11). At the end the Langevin integration is performed (4-8). The interaction between particles has been done by applying the Amber force field [11].

The simulation was performed on a GeForce GT 330M NVIDIA graphics processing unit with 48 CUDA Cores, with a processor clock rate reaching 1265 MHz. By using this GPU, the average execution time for eight simulations where the friction coupling strength for the Langevin integrator and the time constant for the Berendsen thermostat were varied was equal to 9937.61 s. From this it can be stated that in 158.18 minutes are computed 100 picoseconds of MD simulation. The simulation speed was of 910 ps/day and the difference between the reference and the actual temperature did not vary with more than ± 3.5 degrees.

#### B. Simulation on the multi-core systems

The multi-core systems simulation was performed upon the MARTINI coarse-grained force field developed by the molecular dynamics group from the Royal University of Groningen using a system of Martini Water [12]. The MARTINI coarse-grained force field reproduces thermodynamic properties by mapping four heavy atoms to one coarse-grained meta-particle [13, 14]. Due to this mapping, the processing speed is faster because the complexity of computations is reduced.

The transmission of data was done by sending nine messages per process, per time-step, using MPI communication. This mechanics offers scalability, while the sending and receiving of data does not block. The communication is done between processes that belong to the same group (processes can also be part of more than one group as in Fig. 2).

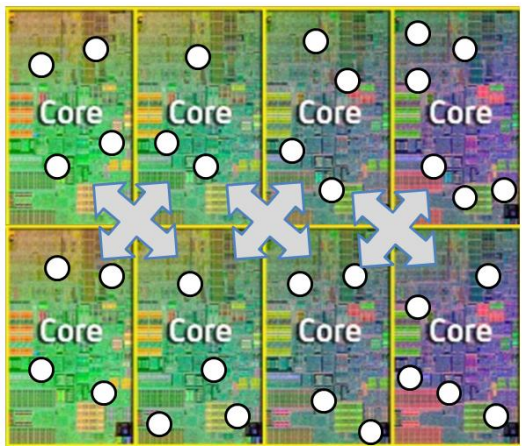

Fig. 2. Multi-core processing on an atomic simulation system

The computers from the Molecular Dynamics Group of the Royal University of Groningen from Netherlands were used for this simulation.

On each processor the following steps took place. Firstly, the Newton interaction forces are computed, followed by the computation of the new velocities (5) and new positions (2). The positions are updated according to the constraints. The compensation of the freedom degrees lost because of constraints is done by generating the Maxwellian velocities for all the particles (12-14).

In our simulation the correction of the degrees of freedom was performed using DPD-ISO (15-19). The coordinates are updated for the $3 * (N - m)$ constraints (20) and the velocities are renewed (21). After this step, the velocities are communicated via MPI.

For the coarse-grained system of 6000 molecules the runtime speed is presented in Table 1, while the average temperature of the system was of 300.321 K.

TABLE I. SIMULATION SPEED FOR VARIOUS NUMBERS OF PROCESSORS

| Total number of processors | Simulation speed [ps/day] |
|---|---|
| 1 | 8197 |
| 2 | 17252 |
| 3 | 23900 |
| 4 | 30084 |
| 5 | 22389 |
| 6 | 25615 |
| 7 | 28430 |
| 8 | 24552 |

*C. Advantages and disadvantages*

The MD simulations on which OpenMM algorithm is run on the graphics processing unit through the CUDA high-level language and the OpenCL platform that permits the kernels to run on the GPU has the average simulation speed equal to 910 ps/day.

For the MD simulations on multi-core systems based on GROMACS and MPI, the use of MPI assures data communication with a constant number of nine messages per process. By parallelizing on multi-core systems the scalability is linear with the number of processors after which, depending on the size of the molecular system, a plateau is reached: earlier for smaller systems or later for larger systems.

The runtime peek from the simulation run on CUDA was obtained for a total number of processors equal to four, the simulation speed being 3.67 times faster in comparison to the use of one processor. It can be stated that increasing the number of cores, speeds up also the execution and this is a combined effect given by the speed up given by the graphical unit combined with the one due to the use of more cores.

For larger molecular systems (to the order of ten thousands of particle) because not all the coordinates and velocities of the molecular system can be loaded on the graphical card there is no gain of using GPU as compared to runs on just CPU's, because the overhead due to the communication between CPU and GPU's card overrun the speed-up obtained by the parallelization on GPU.

Both system variants are thread oriented, supporting parallelization and data communication. The thermostats applied in the two systems provided a system temperature that was very close to the one taken as reference, better results being obtained in the case of DPD-ISO.

In comparison to the CPU that has several cores, the GPU has much more cores, but each one of them runs slower than a CPU core. As we have used them in our molecular dynamics simulation, GPU cores are more appropriate for intensive processing, because for large MD systems there is no gain of using graphical cards for speed-ups (see our earlier comments).

IV. CONCLUSION

The two performed simulations maintained the temperature of the system close to the one taken as reference, as intended, but the simulation speed differed. The simulation on a multi-core system was 33 times faster than the one performed on the graphics processing unit where the number of processors was equal to four.

For a multi-core system where MPI is used the scalability increases as the number of atoms becomes larger. When combining the GPU processing with the one on the CPU, the intensive computational parts of the simulation run on the GPU, while the remained code runs on the CPU. A CPU has just a number of cores which use sequential serial processing, while the GPU applies parallelism, having much more cores that can handle several tasks simultaneously. Moreover, a multi core CPU can provide parallelism for diversity of tasks, sufficient to transform the GPU programming into a superfluous concern, as in the simulation case of GROMACS and MPI system-of-systems.

In conclusion, parallel simulations on multi-core systems are more efficient than the simulations on the graphics processing unit due to the fast processing time, but the costs are greater compared to the later one.

As future work we intend to simulate other types of combined thermostats by applying the PAR and PERP dissipative particle dynamics on multi-core systems, where in the case of PAR the friction is parallel to the radius between the pair of atoms and PERP, when the friction is perpendicular to the radius of the pair of atoms. Another interesting development for future work is the use of MD simulations for music generation in the sense of using the movement of particles for music creation.